\documentclass[twocolumn,showpacs,preprintnumbers,amsmath,amssymb]{revtex4}

\usepackage{graphicx}
\usepackage{dcolumn}

\usepackage{bm}

\begin{document}

\title{Plane symmetric thin-shell wormholes: solutions and stability}
\author{Jos\'e P. S. Lemos}
\email{lemos@fisica.ist.utl.pt} \affiliation{Centro
Multidisciplinar de {}Astrof\'{\i}sica - CENTRA \\
Departamento de F\'{\i}sica, Instituto Superior T\'ecnico - IST,\\
Universidade T\'ecnica de Lisboa - UTL,\\
Avenida Rovisco Pais 1, 1049-001 Lisboa, Portugal}
\author{Francisco S. N. Lobo}
\email{francisco.lobo@port.ac.uk} \affiliation{Institute of
Gravitation \& Cosmology, University of Portsmouth, Portsmouth PO1
2EG, UK} \affiliation{Centro de Astronomia e Astrof\'{\i}sica da
Universidade de Lisboa, Campo Grande, Edif\'{i}cio C8, 1749-016
Lisboa, Portugal}

\date{\today}

\begin{abstract}

Using the cut-and-paste procedure, we construct static and dynamic
plane symmetric wormholes by surgically grafting together two
spacetimes of plane symmetric vacuum solutions with a negative
cosmological constant. These plane symmetric wormholes can be
interpreted as domain walls connecting different universes, having
planar topology, and upon compactification of one or two
coordinates, cylindrical topology or toroidal topology,
respectively. A stability analysis is carried out for the dynamic
case by taking into account specific equations of state, and a
linearized stability analysis around static solutions is also
explored. It is found that thin shell wormholes made of a dark
energy fluid or of a cosmological constant fluid are stable, while
thin shell wormholes made of phantom energy are unstable.

\end{abstract}


\maketitle

\section{Introduction}

There has been much interest in traversable wormholes since the
important Morris-Thorne article \cite{Morris}. It was found that
these geometries, which act as tunnels from one region of
spacetime to another, possess a peculiar property, namely exotic
matter, involving a stress-energy tensor that violates the null
energy condition \cite{Morris,MTY,Visser}, see also \cite{LLQ} for
a review and references therein. In fact, traversable wormholes
violate all of the pointwise energy conditions and averaged energy
conditions (see, e.g., \cite{Lobo:2007zb}). As the violation of
the energy conditions is a particularly problematic issue
\cite{Barcelo}, it is useful to minimize the usage of exotic
matter. Recently, Visser  et al \cite{Kar1,Kar2}, noting the fact
that the energy conditions do not actually quantify the ``total
amount'' of energy condition violating matter, developed a
suitable measure for quantifying this notion by introducing a
``volume integral quantifier''. Although the null energy and
averaged null energy conditions are always violated for wormhole
spacetimes, they considered specific examples of spacetime
geometries containing wormholes that are supported by arbitrarily
small quantities of averaged null energy condition violating
matter.

Another elegant way of minimizing the usage of exotic matter is to
construct a simple class of wormhole solutions using the cut and
paste technique \cite{visser1,visser2} (see also \cite{Visser}),
in which the exotic matter is concentrated at the wormhole throat,
i.e., a thin shell wormhole solution. The surface stresses of the
exotic matter were determined by invoking the Darmois-Israel
formalism \cite{Israel}. These thin-shell wormholes are extremely
useful as one may apply a stability analysis for the dynamical
cases, either by choosing specific surface equations of state
\cite{eqstate1}, or by considering a linearized stability analysis
around a static solution
\cite{Brady,Poisson,Ishak,Eiroa,Lobolinear,Lobolinear2}, in which
a parametrization of the stability of equilibrium is defined, so
that one does not have to specify a surface equation of state. The
dynamical analysis in \cite{Brady,Poisson,Ishak} was generalized
by considering solutions with electrical charge \cite{Eiroa},
solutions in the presence of a cosmological constant
\cite{Lobolinear}, and spherically symmetric dynamical solutions
\cite{Lobolinear2}. More recently, by using the cut and paste
technique, several solutions were analyzed, such as, the dynamics
of non rotating cylindrical thin-shell wormholes
\cite{Eiroa:2004at}, charged thin-shell Lorentzian wormholes in
dilaton gravity \cite{Eiroa:2005pc}, five dimensional thin-shell
wormholes in Einstein-Maxwell theory with a Gauss-Bonnet term
\cite{Thibeault:2005ha}, solutions in higher dimensional
Einstein-Maxwell theory \cite{Rahaman:2006vg}, thin-shell
wormholes associated with global cosmic strings
\cite{Bejarano:2006uj}, solutions in heterotic string theory
\cite{Rahaman:2006xb}, spherical thin-shell wormholes supported by
a Chaplygin gas \cite{Eiroa:2007qz}, a new class of thin-shell
wormhole by surgically grafting two black hole solutions localized
on a three brane in five dimensional gravity in the
Randall-Sundrum scenario \cite{Rahaman:2007bf}, and spherically
symmetric thin-shell wormholes in a string cloud background in
(3+1)-dimensional spacetime were also analyzed
\cite{Richarte:2007bx}. Other wormhole solutions have been
analyzed in \cite{Lobo:2006ue,Sushkov,Lobo-phantom,Eiroa:2008ky}.

One can go beyond thin shell solutions, and construct wormhole
solutions by matching an interior wormhole to an exterior vacuum
solution, at a junction surface. In particular, a thin shell
around a traversable wormhole, with generic surface stresses  was
analyzed in \cite{Lobo-CQG}, a particular case of this, a wormhole
with a zero surface energy density and constant redshift function,
having been studied in \cite{LLQ}. A similar analysis for the
plane symmetric case, with a negative cosmological constant, was
done in \cite{LL}. The plane symmetric traversable wormhole is a
natural extension of topological black hole solutions with a
negative cosmological constant
\cite{lemos,zanchin,lemosreview,Lemos4}, upon addition of exotic
matter. These plane symmetric wormholes can be interpreted as
domain walls connecting different universes, having planar
topology, and upon compactification of one or two coordinates,
cylindrical topology or toroidal topology, respectively. The
construction of these wormholes does not alter the topology of the
background spacetime (i.e., spacetime is not multiply-connected),
so that these solutions can instead be considered domain walls.
Thus, in general, these wormhole solutions do not allow time
travel.

From recent astronomical observations, it seems that we presently
live in a world with a positive cosmological constant,
$\Lambda>0$. However, a spacetime with $\Lambda<0$ is also of
significant relevance, since it allows a consistent physical
interpretation in the context of supergravity and superstring
theories \cite{lemosreview}. Indeed, one can enlarge general
relativity into a gauged extended supergravity theory, in which
the vacuum state has an energy density given by
$\Lambda=-3g^2/(4\pi pG)$, where $g$ is the coupling constant of
the theory, and $G$ is the gravitational constant. Thus, the
vacuum of the theory is described by an anti-de Sitter spacetime.
It is important to emphasize that if these theories are correct,
they imply that the anti-de Sitter spacetime should be considered
as a symmetric phase of the theory, although it must have been
broken, since we do not presently live in a universe with
$\Lambda<0$. In addition, note that anti-de Sitter spacetimes have
other interesting features: (i) they are one of the rare
gravitational backgrounds yielding a consistent interaction with
massless higher spins, (ii) they permit a consistent theory of
strings in any dimension, (iii) they allow valid definitions for
the mass, angular momentum and charge, and (iv) it has been
conjectured that they has a direct correspondence with conformal
field theory on the boundary of that space, the AdS-CFT
conjecture, see, e.g., \cite{Aharony:1999ti}. Moreover, even with
its preference to negative cosmological constant scenarios, string
theory can although in a contrived way, produce a landscape of
positive cosmological constant universes \cite{kklt}, indicating
perhaps that one can transit bewteen both signs of the
cosmological constant. So, it is certainly of interest to study
astrophysical structures that appear within a positive
cosmological constant scenario, as well as to study micro
structures that may appear within a negative cosmological constant
scenario. Here we analyze wormhole structures within a negative
cosmological context. Within this context other structures, such
as black holes with spherical, toroidal, and hyperbolical
topologies, have been analyzed (see \cite{lemosreview} for a
review).

The paper is organized as follows. By using the cut-and-paste
procedure for a thin shell, we construct static and dynamic
wormholes by surgically grafting together two spacetimes of plane
symmetric vacuum solutions, with planar, cylindrical or toroidal
topologies, with a negative cosmological constant. This analysis
is displayed in Section \ref{Sec:II}. We shall also consider the
specific case of static wormhole solutions, and consider several
equations of state. In Section \ref{Sec:III}, we consider a
dynamical stability analysis. In particular, a stability analysis
is carried out for the dynamic case by taking into account
specific surface equations of state, and a linearized stability
analysis around static solutions is also further explored.
Finally, in Section \ref{Conclusion}, we conclude.

\section{Black membrane surgery and static wormholes}\label{Sec:II}

\subsection{Cut and paste technique}

\subsubsection{General considerations, general equation of state}

The planar black hole or a black membrane metric is given by
\cite{lemos,zanchin,lemosreview,Lemos4}
\begin{eqnarray}
ds^2&=&-\left(\alpha^2r^2-\frac{M}{\alpha r} \right)
\,dt^2+\left(\alpha^2r^2-\frac{M}{\alpha r} \right)^{-1}\,dr^2
     \nonumber   \\
&&+\alpha^2\,r^2\,(dx ^2+dy^2) \label{planarBH}\,,
\end{eqnarray}
where $\alpha$ is the inverse of the characteristic length of the
system, which here we adopt as given by the negative cosmological
constant, i.e., we put $\alpha^2=-\Lambda/3$. The ranges of $t$
and $r$ are $-\infty <t<+\infty$ and $0\leq r<+\infty$.  The range
of the coordinates $x$ and $y$ determine the topology of the plane
symmetric metric. For the planar case, the topology of the
two-dimensional space, $t={\rm constant}$ and $r={\rm constant}$,
is $R^2$, with coordinate range $-\infty <x<+\infty$ and $-\infty
<y<+\infty$. For the cylindrical case the topology is $R\times
S^1$, with $-\infty <x<+\infty$ and $0 \leq \alpha\, y<2\pi$. For
the toroidal case the topology is $S^1\times S^1$ (i.e., the torus
$T\,^2$), with $0 \leq \alpha\, x<2\pi$ and $0 \leq \alpha\,
y<2\pi$. A scalar polynomial singularity occurs at $r=0$, which
can be demonstrated through the Weyl scalar, given by
$C^{{\mu}{\nu}{\alpha}{\beta}} C_{{\mu}{\nu}{\alpha}{\beta}}
={12M^2}/{(\alpha^2\,r^6)}$, where Greek indices are spacetime
indices, $\mu=t,r,x,y$. It can also be found through the
Kretschmann scalar, given by $ R^{{\mu}{\nu}{\alpha}{\beta}}
R_{{\mu}{\nu}{\alpha}{\beta}}
=24\alpha^4+12{M^2}/{(\alpha^2\,r^6)}$, but since the first term
is innocuous, the Weyl scalar suffices. In particular, we shall
consider the planar topology case, and the analysis considered in
this work applies equally well for solutions with toroidal and
cylindrical topologies. Note that considering a positive value for
$M$, an event horizon occurs at $r_h=M^{1/3}/\alpha$, consequently
resulting in a planar black hole solution, or black membrane.
Considering a negative value for $M$, we verify that no event
horizons occur, implying the existence of a singular naked
membrane. A remark is in order here: cylindrically thin-shell
wormholes were analyzed in \cite{Eiroa:2004at}, which are
different to the planar, cylindrical or toroidal topological
solutions we analyze, as for our case infinity carries the same
topology as the throat

An interesting wormhole solution consists in applying the
cut-and-paste construction. We consider two copies of the black
membrane or planar black hole solution, Eq. (\ref{planarBH}),
removing from each spacetime the four-dimensional regions
described by
\begin{equation}
\Omega^{\pm}\equiv \left \{r^{\pm}\leq a| \,a > M^{1/3}/\alpha
\right \} \,,
\end{equation}
where $a$ is a constant. The condition $a > M^{1/3}/\alpha$ is
important to avoid the presence of an event horizon. The removal
of these two regions results in two manifolds, geodesically
incomplete, with boundaries given by the following timelike
hypersurfaces
\begin{equation}
\partial \Omega^{\pm}\equiv \left \{r^{\pm}= a| \,a > M^{1/3}/\alpha
\right \} \,.
\end{equation}
We now identify these two timelike hypersurfaces, $\partial
\Omega^{+}= \partial \Omega^{-}\equiv \partial \Omega$, which
results in a manifold, now geodesically complete, with two regions
connected by a wormhole. The throat of the wormhole, i.e., the
junction surface, is situated at $\partial \Omega$, and may be
viewed as behaving like a domain wall connecting two universes.
One can then write the intrinsic metric to $\partial \Omega$ as
\begin{equation}
ds^2_{\partial \Omega}=-d\tau^2 + a^2(\tau) \,\left(dx
^2+dy^2\right) \,,
\end{equation}
where $\tau$ is the proper time on $\partial \Omega$.

In general terms, the position of the junction surface is given by
$x^{\mu}(\tau,x,y)$, and the respective $4$-velocity is
\begin{eqnarray}
u^{\mu}_{\pm}\equiv \frac{dx^\mu}{d\tau}\,,
\end{eqnarray}
where $\tau$ is the proper time on the surface. We shall use the
Darmois-Israel formalism to determine the surface stresses at the
junction boundary \cite{Israel}. The intrinsic surface
stress-energy tensor, $S_{ij}$, is given by the Lanczos equations
in the form $S^{i}_{\;j}=-\frac{1}{8\pi}\,(\kappa
^{i}_{\;\,j}-\delta ^{i}_{\;j}\kappa ^{k}_{\;\,k})$, where Latin
indices are intrinsic surface indices which run through
$i=\tau,x,y$. For notational convenience, the discontinuity in the
second fundamental form or extrinsic curvatures is given by
$\kappa_{ij}=K_{ij}^{+}-K_{ij}^{-}$. The second fundamental form
is defined as
\begin{eqnarray}
K_{ij}^{\pm}&=&\frac{\partial x^{\alpha}}{\partial \xi ^{i}} \,
\frac{\partial x^{\beta}}{\partial \xi
^{j}}\,\nabla_{\alpha}^{\pm}\,n_{\beta}
      \nonumber    \\
&=&-n_{\gamma} \left(\frac{\partial ^2 x^{\gamma}}{\partial \xi
^{i}\,\partial \xi ^{j}}+\Gamma ^{\gamma \pm}_{\;\;\alpha
\beta}\;\frac{\partial x^{\alpha}}{\partial \xi ^{i}} \,
\frac{\partial x^{\beta}}{\partial \xi ^{j}} \right)
   \label{defextrinsiccurvature}     \,,
\end{eqnarray}
where $n_{\gamma}$ is the unit $4$-normal to $\partial \Omega$,
and the superscripts $\pm$ correspond to the exterior and interior
spacetimes, respectively. The parametric equation for $\partial
\Omega$ is given by $f(x^{\mu}(\xi ^{i}))=0$. The unit $4$-normal
to $\partial \Omega$ is given by
\begin{equation}
n_{\mu}=\pm \left|\, g^{\alpha \beta}\, \frac{\partial f}{\partial
x^{\alpha}}\, \frac{\partial f}{\partial x^{\beta}} \right|^{-1/2}
\,\frac{\partial f}{\partial x^{\mu}} \,,
         \label{defnormal}
\end{equation}
with $n^{\mu}\,n_{\mu}=+1$.

Applying to our particular case, the position of the junction
surface is given by $x^{\mu}(\tau,x,y)=(t(\tau),a(\tau),x,y)$, and
the respective $4$-velocity is
\begin{eqnarray}
u^{\mu}_{\pm}\equiv \left(\frac{dt}{d\tau},\frac{da}{d\tau},0,0
\right)= \left(\frac{\sqrt{\alpha^2r^2-\frac{M}{\alpha
r}+\dot{a}^2}}{\alpha^2r^2-\frac{M}{\alpha r}}\;, \dot{a},0,0
\right),
\end{eqnarray}
where the overdot denotes a derivative with respect to $\tau$,
which is defined as the proper time on $\partial \Omega$. The unit
normal to the junction surface may be determined by Eq.
(\ref{defnormal}) or by the contractions, $u^{\mu}n_{\mu}=0$ and
$n^{\mu}n_{\mu}=+1$, and is given by
\begin{eqnarray}
{n_{\mu}}_{\pm}= \left(-\dot{a},
\frac{\sqrt{\alpha^2r^2-\frac{M}{\alpha r}+\dot{a}^2}}
{\alpha^2r^2-\frac{M}{\alpha r}},0,0 \right)\,. \label{normal}
\end{eqnarray}
Considerable simplifications occur due to plane symmetry, namely
$\kappa ^{i}_{\;j}={\rm diag} \left(\kappa
^{\tau}_{\;\,\tau},\kappa ^{x}_{\;\,x},\kappa
^{x}_{\;\,x}\right)$. Thus, the surface stress-energy tensor may
be written in terms of the surface energy density, $\sigma$, and
the surface pressure, ${\cal P}$, as $S^{i}_{\;j}={\rm
diag}(-\sigma,{\cal P},{\cal P})$, which taking into account the
Lanczos equations, reduce to
\begin{eqnarray}
\sigma &=&-\frac{1}{4\pi}\kappa ^{x}_{\;\,x} \,,   \label{sigma} \\
{\cal P} &=&\frac{1}{8\pi}(\kappa ^{\tau}_{\;\,\tau}+\kappa
^{x}_{\;\,x})    \label{surfacepressure}  \,.
\end{eqnarray}
This simplifies the determination of the surface stress-energy
tensor to that of the calculation of the non-trivial components of
the extrinsic curvature, or the second fundamental form.
Therefore, taking into account the metric of Eq. (\ref{planarBH})
and the definition of the second fundamental form, Eq.
(\ref{defextrinsiccurvature}), we have
\begin{eqnarray}
K ^{\tau \;\pm}_{\;\,\tau}&=&\pm\frac{\alpha ^2 a+\frac{M}{2\alpha
a^2}+\ddot{a}}
{\sqrt{\alpha ^2 a^2- \frac{M}{\alpha a}}+\dot{a}^2}  \;, \\
K^{x\;\pm}_{\;\,x}&=&\pm \frac{1}{a}\sqrt{\alpha ^2
a^2-\frac{M}{\alpha a}+\dot{a}^2}  \;.
\end{eqnarray}
From Eqs. (\ref{sigma})-(\ref{surfacepressure}), we deduce the
surface stress-energy components given by
\begin{eqnarray}
\sigma&=&-\frac{1}{2\pi a} \; \sqrt{\alpha ^2
a^2-\frac{M}{\alpha a}+\dot{a}^2}  \label{dynsurfaceenergy} \,,\\
{\cal P}&=&\frac{1}{4\pi a} \; \frac{2\alpha^2
a^2-\frac{M}{2\alpha a}+\dot{a}^2+a\ddot{a}}{\sqrt{\alpha^2
a^2-\frac{M}{\alpha a}+\dot{a}^2}} \label{dynsurfacepressure} \,,
\end{eqnarray}
respectively. Obviously $\sigma$ and $\cal P$ obey the
conservation equation
\begin{equation}
\left(\sigma a^2\right)\dot{}+{\cal P}\left(a^2\right)\dot{}=0 \,.
\label{conservationequation}
\end{equation}

In order to be able to solve the two nontrivial equations
(\ref{dynsurfaceenergy})-(\ref{dynsurfacepressure}) for
$\sigma(\tau)$ and $a(\tau)$, one needs to specify an equation of
state. Here we impose a cold equation of state \begin{equation}
{\cal P}={\cal P}(\sigma)\,, \label{generalstateequation}
\end{equation} which follows if the temperature is zero, $T=0$, in
a general equation of state of the form ${\cal P}={\cal
P}(\Sigma,T)$, $\sigma=\sigma(\Sigma,T)$, where $\Sigma$ is the
baryonic mass density.  In our study we will make the analysis
initially with a generic cold equation of state of the form
(\ref{generalstateequation}), i.e., without invoking any specific
equation of state, see \cite{Poisson}, and then particularizing to
a specific cold equation of state, see \cite{Visser}.

\subsubsection{Specific equation of state: dark energy}

Besides doing the analysis for a general equation of state of the
form ${\cal P}={\cal P}(\sigma)$, as in Eq.
(\ref{generalstateequation}), we want to study specific cases by
giving an equation of state. We resort in the following to a
particularly interesting equation of state, namely, an equation of
state  analogous to the dark energy equation of state considered
in cosmology, i.e.,
\begin{equation}
{\cal P}=\omega\sigma\,,\quad {\rm with}\quad\omega<0\,.
\label{eos}
\end{equation}
Such a dark energy fluid can be divided into three cases: a normal
dark energy fluid when $-1<\omega<0$, a cosmological constant
fluid when $\omega=-1$, and a phantom energy fluid when
$\omega<-1$. All of these fluids are possible candidates
responsible for the accelerated expansion of the Universe. But in
addition to its cosmological interest, the equation of state
(\ref{eos}) can also be used for domain walls and wormholes,
connecting opposite spacetime regions.

\subsection{Static wormholes}

\subsubsection{General considerations, general equation of state}

The case of a static wormhole is particularly simple, yet it
provides interestingly enough results. Equations
(\ref{dynsurfaceenergy})-(\ref{dynsurfacepressure}) reduce to
\begin{eqnarray}
\sigma&=&-\frac{1}{2\pi a} \; \sqrt{\alpha ^2
a^2-\frac{M}{\alpha a}}   \label{membranesigma}   \,,\\
{\cal P}&=&\frac{1}{2\pi a} \; \frac{\alpha^2 a^2-\frac{M}{4\alpha
a}}{\sqrt{\alpha^2 a^2-\frac{M}{\alpha a}}}
\label{membranepressure}        \,.
\end{eqnarray}
Since there is a generic equation of state of the type given in
Eq. (\ref{generalstateequation}), the constants $M$, $a$, and
$\alpha$ are inter-related.

The surface energy density in (\ref{membranesigma}) is negative,
implying the violation of the weak and dominant energy conditions.
The surface pressure is always positive. The null energy condition
is verified if $\sigma+{\cal P} >0$ is satisfied. Thus, taking
into account the relationship
\begin{eqnarray}
\sigma+{\cal P}&=&\frac{1}{4\pi a} \; \frac{\frac{3M}{2\alpha
a}}{\sqrt{\alpha^2 a^2-\frac{M}{\alpha a}}} \,,
       \label{sigma+P}
\end{eqnarray}
we see that the null energy condition is verified only for $M>0$,
and violated for $M<0$.  The strong energy condition is satisfied
if $\sigma+{\cal P} >0$ and $\sigma+2{\cal P} >0$, and by
continuity implies the null energy condition. Using the condition
\begin{eqnarray}
\sigma+2{\cal P}&=&\frac{1}{2\pi a} \;\frac{\alpha^2
a^2+\frac{M}{2\alpha a}}{\sqrt{\alpha^2 a^2-\frac{M}{\alpha a}}}
\,,
            \label{sigma+2P}
\end{eqnarray}
we verify that the strong energy condition is readily satisfied
for $M>0$.

It is also of interest to analyze the attractive or repulsive
character \cite{Lobo:2006ue} of this plane symmetric traversable
wormhole. The four-velocity of a static observer is
$u^{\mu}=dx^{\mu}/{d\tau} =(u^{\,t},0,0,0)=(1/\sqrt{\alpha^2
r^2-M/\alpha r}\,,0,0,0)$. The observer's four-acceleration is
$a^{\mu} =u^{\mu}{}_{;\nu}\,u^{\nu}$, so that taking into account
Eq.~(\ref{planarBH}), the only non-zero component is given by
\begin{eqnarray}
a^r = \Gamma^r_{tt}\,\left(\frac{dt}{d\tau}\right)^2
    =\alpha^2 r+\frac{M}{2\alpha r^2}\,.         \label{radial-acc}
\end{eqnarray}
From the geodesic equation, a radially moving test particle which
starts from rest initially has the equation of motion
\begin{equation}
\frac{d^{\,2}r}{d{\tau}^2}=-\Gamma^r_{tt}\,
\left(\frac{dt}{d\tau}\right)^2 =-a^r\,.
      \label{radial-accel}
\end{equation}
Therefore, $a^r$ is the radial component of proper acceleration
that an observer must maintain in order to remain at rest at
constant $(r,x,y)$. It is interesting to note that a wormhole is
``attractive'' if $a^r>0$, i.e., observers must maintain an
outward-directed radial acceleration to keep from being pulled
into the wormhole; and ``repulsive'' if $a^r<0$, i.e., observers
must maintain an inward-directed radial acceleration to avoid
being pushed away from the wormhole. Note that for $M>0$, the
wormhole is attractive. For $M<0$, the wormhole is repulsive for
the values $r<\sqrt[3]{|M|/2}/\alpha$, and attractive for
$r>\sqrt[3]{|M|/2}/\alpha$. In particular, for
$r=\sqrt[3]{|M|/2}/\alpha$ static observers are geodesic.

\subsubsection{Specific equation of state: dark energy}

To justify the equation of state (\ref{eos}) let us first try a
well known case, the surface stress-energy tensor without trace.
As we will show now, this is no good. The traceless surface
stress-energy tensor, $S^{i}_{\;\,i}=0$, i.e., $-\sigma +2{\cal
P}=0$, could be a case of particular interest, since the Casimir
effect with a massless field gives a stress-energy tensor of this
type. This effect is theoretically invoked to provide exotic
matter to the system considered at hand. From $-\sigma +2{\cal
P}=0$ we deduce $\alpha^2 a^2=M/(2\alpha a)$, which is inside the
event horizon of the planar black hole, or the black membrane, and
so is no good.  As in the spherical symmetric case \cite{Visser},
there is no solution of the Einstein field equations because
$\sigma$ and ${\cal P}$ are imaginary, as one may readily verify
from Eqs. (\ref{membranesigma})-(\ref{membranepressure}).  So, one
has to resort to the equation of state given in (\ref{eos}), i.e.,
${\cal P}=\omega \sigma$, with $\omega<0$.  That $\omega$ has to
be less than zero can be seen directly from equations
(\ref{membranesigma})-(\ref{membranepressure}). Indeed, as the
surface energy density is always negative, $\sigma<0$, and the
tangential surface pressure is always positive, ${\cal P}>0$, then
the equation of state (\ref{eos}) imposes the condition
$\omega<0$, i.e., a dark energy equation of state.  As mentioned
above, one can subdivide the $\omega<0$ fluid further into a
normal dark energy fluid ($-1<\omega<0$), a cosmological constant
fluid ($\omega=-1$), and a phantom energy fluid ($\omega<-1$) This
type of dark energy fluids are considered in cosmology, as well as
in domain wall and wormhole building.  Indeed, for the latter, the
particular case of $\omega=-1$ reduces to $\sigma +{\cal P}=0$,
which corresponds to the classical membrane, this being described
by the three-dimensional generalization of the Nambu-Goto action,
and being the simplest domain wall one may construct
\cite{Visser}. From Eq. (\ref{sigma+P}), one finds that this case
is only valid when $M=0$, thus from Eqs.
(\ref{membranesigma})-(\ref{membranepressure}) the surface energy
density and surface pressure are given by $\sigma=-{\cal
P}=-\frac{\alpha}{2\pi}$.  Another interesting example, is when
$\omega<-1$, where it has been shown that spherically symmetric
traversable wormholes can be theoretically supported by phantom
energy \cite{Sushkov,Lobo-phantom}.

\begin{figure}[h]
  \centering
  \includegraphics[width=3.4in]{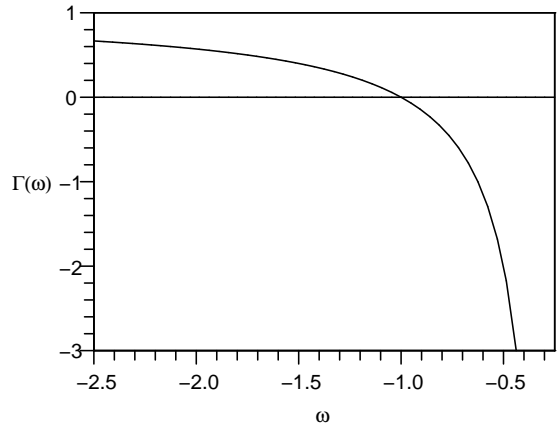}
  \caption{We have defined $\Gamma(\omega)=(1+\omega)/(1/4+\omega)$.
  We verify that if $M>0$ then $\Gamma(\omega)>0$, and
$\omega$ has the range $\omega<-1$ and $-1/4<\omega<0$. However,
the latter range in ruled out by imposing the additional no
horizon condition $M<\alpha^3a^3$, so that the $M>0$ case is
restricted to $-\infty<\omega<-1$. For $M=0$ one has $\omega=-1$.
If $M<0$ then $\Gamma(\omega)<0$, and $\omega$ is restricted
within the range $-1<\omega<-1/4$. See the text for details.}
\label{Gamma-range}
\end{figure}

Now, using Eqs. (\ref{membranesigma})-(\ref{membranepressure}), we
obtain the following relationship
\begin{equation}
M=\Gamma(\omega)\,\alpha^3a^3\,, \label{Gamma}
\end{equation}
where
\begin{equation}
\Gamma(\omega)\equiv\frac{1+\omega}{1/4+\omega} \,.
\end{equation}
Note that if $M>0$ then $\omega$ has the range $-\infty<\omega<-1$
and $-1/4<\omega<0$.  One should impose the additional no horizon
condition, $M<\alpha^3a^3$, since if the condition is not
satisfied there are no static solutions, and moreover the
wormholes would be inside their own gravitational radius, a
situation which is of no interest here.  Thus, imposing the
additional no horizon condition one gets from Eq. (\ref{Gamma})
that $\Gamma(\omega)<1$, which means the $M>0$ case is restricted
to $-\infty<\omega<-1$. If $M=0$, i.e., the fluid is the spacetime
background fluid with a negative cosmological constant, it
naturally gives the corresponding equation of state, i.e.,
$\omega=-1$. If $M<0$ then $\omega$ is restricted to the range
$-1<\omega<-1/4$. Thus the full range of $\omega$ is
$-\infty<\omega<-1/4$, see Fig.  \ref{Gamma-range}.

\section{Dynamical stability analysis}\label{Sec:III}

\subsection{Stability analysis through matter fields}\label{stability}

\subsubsection{General considerations with a general equation of state}

To analyze the dynamics of the wormhole, we consider that the
throat is a function of the proper time, as measured by an
observer comoving with the throat \cite{Visser, Poisson, Brady}.
Assume that the position of the throat is given by
$x^{\mu}(\tau,x,y)=(t(\tau),a(\tau),x,y)$. As mentioned, Eqs.
(\ref{dynsurfaceenergy})-(\ref{dynsurfacepressure}) imply the
energy conservation equation (\ref{conservationequation}), which
in turn can be put in the form
\begin{equation}
\dot{\sigma}=-2 \left(\sigma +{\cal P} \right)\frac{\dot{a}}{a}
    \label{conservationofenergy}    \,.
\end{equation}
Equation (\ref{conservationofenergy}) may also be rewritten as
\begin{equation}
\frac{da}{a}=-\frac{1}{2}\;\frac{d\sigma}{\sigma+{\cal P}}  \,.
     \label{conservationofenergy2}
\end{equation}
Taking into account an equation of state of the form given in Eq.
(\ref{generalstateequation}), i.e., ${\cal P}={\cal P}(\sigma)$,
Eq. (\ref{conservationofenergy2}), can be integrated to yield
\begin{equation}
\ln(a)=-\frac{1}{2}\int \frac{d\sigma}{\sigma +{\cal P}(\sigma)}
\,.     \label{integral}
\end{equation}
This result may be formally inverted to provide
$\sigma=\sigma(a)$. Equation (\ref{dynsurfaceenergy}) may be
recast into the following form
\begin{equation}
\dot{a}^2-\left[(2\pi \sigma(a))^2-\alpha^2
\right]a^2-\frac{M}{\alpha a}=0
    \label{totalenergy}   \,,
\end{equation}
which may also be written as $\dot{a}^2=-V(a)$, where the
potential is defined as
\begin{equation}
V(a)=-\left[(2\pi \sigma)^2-\alpha^2 \right]a^2-\frac{M}{\alpha a}
\,.
   \label{potential}
\end{equation}
We sketch an analysis of this potential by dividing it into three
cases. (i) When $(2\pi \sigma(a))^2>\alpha^2$, from Eq.
(\ref{totalenergy}), we can divide into two cases $M\geq 0$ and
$M<0$.  For $M\geq0$ we verify that the wormhole is dynamically
unstable. Depending on the initial conditions, this wormhole
solution will either collapse to $a=0$ or explode to $a
\rightarrow \infty$. For $M<0$ the wormhole never collapses, and
depending on the form of $\sigma(a)$ it may or may not explode.
Thus, for $M<0$ the solution may be stable. (ii) When $(2\pi
\sigma(a))^2=\alpha^2$, then Eq. (\ref{totalenergy}) reduces to
$\dot{a}^2-M/\alpha a=0$.  Thus, we can divide also into two cases
$M\geq 0$ and $M<0$ For $M\geq 0$ the differential equation has
the following solutions $ a(\tau)=\frac{\alpha}{4M} \left(\pm 12
M^2\alpha^{-2} \tau \mp 12 M^2\alpha^{-2} C \right)^{2/3} \,, $
where $C$ is a constant related to the initial radius of the
wormhole, $C=\frac{2}{3}\sqrt{\alpha a_0^3/M}$. We verify that if
$\tau \rightarrow \infty$, the wormhole explodes, $a \rightarrow
\infty$. The wormhole collapses to $a\rightarrow 0$ in the proper
time $\tau=C$. In addition to $(2\pi \sigma)^2=\alpha^2$,
considering $M=0$, we have a stable and static wormhole solution,
i.e., $\dot{a}(\tau)=0$, as we verified whilst analyzing the
classical membrane in static wormholes.  For $M<0$ there is no
solution. (iii) When $(2\pi \sigma(a))^2<\alpha^2$, for $M\geq 0$
one has that the wormhole will not explode, but will collapse. For
$M<0$ there is no solution.

Before entering the analysis for the specific dark energy equation
of state (\ref{eos}), we give a simple example of a dynamically
stable solution with a specific choice for the matter field
surface energy density. Note that to be dynamically stable, the
potential is required to be bounded from above and from below. To
achieve this, one may consider the following specific choice for
the surface energy density
$ \sigma(a)=\pm
\frac{1}{2\pi}\left\{\alpha^2-\frac{1}{a^2}\left[k_1(a-k_2)^2-k_3
+\frac{M}{\alpha a}\right]\right\}\,, $
where $k_i>0$ (with $i=1,2,3$) are constants. Substituting this
expression in Eq. (\ref{potential}), provides the following
potential
$ V(a)=k_1(a-k_2)^2-k_3\,, $
which can easily be depicted qualitatively.
The zeroes of $V(a)$ are situated at
$a_{1,2}=k_2\mp\sqrt{k_3/k_1}$, where the lower root
$a_1=k_2-\sqrt{k_3/k_1}$ is required to obey $a_1>M^{1/3}/\alpha$,
in order to avoid a black hole solution.

\subsubsection{Specific equation of state: dark
energy} \label{darkenergy}

\begin{figure}[h]
  \centering
  \includegraphics[width=2.6in]{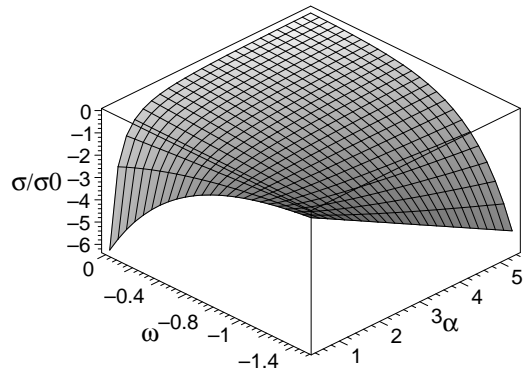}
  \caption{We have defined $\alpha=a/a_0$. As $\omega \rightarrow 0$
and for high values of $\alpha=a/a_0$, then $\sigma \rightarrow
0$. On the other hand,  for decreasing values of the parameter
$\omega$ and for high $\alpha=a/a_0$, then $\sigma \rightarrow
-\infty$. For $\omega \rightarrow 0$ and $\alpha=a/a_0 \rightarrow
0$, then $\sigma \rightarrow -\infty$. See the text for details.}
\label{sigma(omega)}
\end{figure}

Considering the equation of state given in Eq. (\ref{eos}), i.e.,
${\cal P}=\omega \sigma$, from Eq.  (\ref{integral}), we deduce
\begin{equation}
\sigma(a)=\sigma_0\,\left(\frac{a_0}{a}\right)^{2(1+\omega)} \,,
   \label{darksigma}
\end{equation}
where $a_0$ is the initial position of the throat and
$\sigma_0=\sigma(a_0)$. The qualitative behavior is transparent
from Fig. \ref{sigma(omega)}. As $\omega \rightarrow 0$ and for
high values of $\alpha=a/a_0$, then $\sigma \rightarrow 0$. On the
other hand for decreasing values of the parameter $\omega$ and for
high $\alpha=a/a_0$, then $\sigma \rightarrow -\infty$. For
$\omega \rightarrow 0$ and $\alpha=a/a_0 \rightarrow 0$, then
$\sigma \rightarrow -\infty$. Note that considering the equation
of state of the domain wall \cite{Visser, Ishak}, $\sigma +{\cal
P}=0$, from Eq. (\ref{conservationofenergy}) we have
$\dot{\sigma}=0$, i.e., $\sigma ={\rm const}<0$.

\begin{figure}[h]
\centering
\includegraphics[width=3.4in]{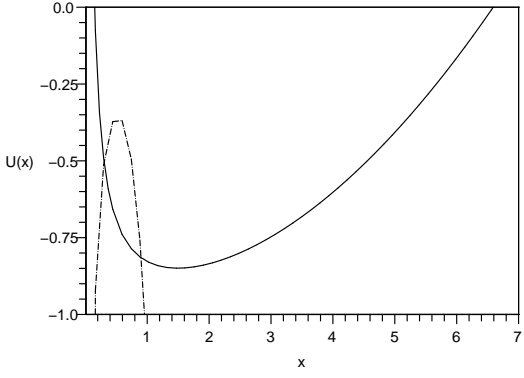}
\caption{Stability analysis for the specific case of a dark energy
equation of state, with $x=a/a_0$. For positive values of
$\bar{M}$, depicted by the dashed curve, we have considered the
specific numerical values of $\bar{M}=0.15$, $\omega=-3/2$ and
$\bar{\alpha}=0.15$. We verify that for this case, the wormhole is
unstable. For $\bar{M}<0$, depicted by the solid curve, one has
stable wormhole solutions. For this case, we have considered the
specific numerical values of $\bar{M}=-0.15$, $\omega=-1/2$ and
$\bar{\alpha}=0.15$. See the text for details.}
\label{plot:darkplot}
\end{figure}

Relative to the stability analysis, by substituting Eq.
(\ref{darksigma}) into the potential, Eq.  (\ref{potential}),
yields the following relationship
\begin{equation}\label{darkpot}
U(a)=-\left[\left(\frac{a}{a_0}\right)^{-4(1+\omega)}
-\bar{\alpha}^2\right]\left(\frac{a}{a_0}\right)^2
-\bar{M}\left(\frac{a_0}{a}\right)\,,
\end{equation}
where the following definitions were considered for notational
simplicity
\begin{equation}
U(a)=\frac{V(a)}{\left(a_0 \bar{\sigma}_0\right)^2}\,, \quad
\bar{\alpha}^2=\frac{\alpha^2}{\bar{\sigma}_0^2}\,, \quad
\bar{M}=\frac{M}{\bar{\alpha}\left(a_0 \bar{\sigma}_0\right)^3}\,,
\end{equation}
with $\bar{\sigma}_0=2\pi\sigma_0$. A stability analysis is
depicted in Fig. \ref{plot:darkplot} for specific numerical
values, which may be considered as representative for the present
stability analysis. For positive values of $\bar{M}$, depicted by
the dashed curve, we have considered the specific values of
$\bar{M}=0.15$, $\omega=-3/2$ and $\bar{\alpha}=0.15$. One
verifies that for this case, the wormhole is unstable. For
$\bar{M}<0$, depicted by the solid curve, we have considered the
specific numerical values of $\bar{M}=-0.15$, $\omega=-1/2$ and
$\bar{\alpha}=0.15$, and one verifies the stability of the
wormhole.

Thus, in summary, for $\bar M>0$, one sees that the wormholes are
unstable, as depicted by the dashed curve in Fig.
\ref{plot:darkplot}.  For $\bar M=0$, not depicted, the wormhole
is static and neutrally stable. For $\bar M<0$, which is depicted
by the solid curve in Fig. \ref{plot:darkplot}, the wormhole is
stable against expansion and collapse, it will oscillate between a
maximum value and a minimum
value of $a$. %

\subsection{Linearized stability analysis}

\subsubsection{General considerations with a general equation of state}
\label{gencon}

An alternative to the partial stability analysis of Section
\ref{stability} is to consider a linear perturbation around a
static solution with radius $a_0$. The respective static values of
the surface energy density and the surface pressure are given by
Eqs. (\ref{membranesigma})-(\ref{membranepressure}).  To know
whether the equilibrium solution is stable or not, one must
analyze the shell's equation of motion near the equilibrium
solution. This means that in order to solve this equation one has
to know the equation of state of the shell's matter.

\begin{figure}[t]
\centering
\includegraphics[width=2.3in]{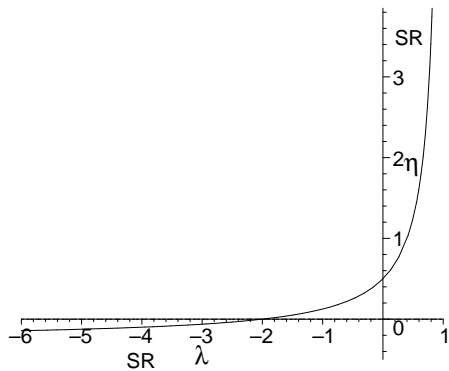}
\caption{The regions of stability in a plot of
$\eta_0\times\lambda$. Here $\lambda$ is defined as
$\lambda=\frac{M}{(\alpha a_0)^3}$. The range of $\lambda$ is
$-\infty<\lambda<1$.  SR in the plot means stability region. For
$0<\lambda<1$ (i.e., positive masses, $M>0$) the stability region
is above the curve shown and to the right of the axis $\lambda=0$.
For $\lambda=0$ (i.e., zero mass, $M=0$) the stability region is
the whole $\eta_0$ axis, i.e., the axis $\lambda=0$ itself. For
$\lambda<0$ (i.e., negative masses, $M<0$) the stability region is
below the curve shown and to the left of the axis $\lambda=0$. }
\label{fig:stable1}
\end{figure}

In this subsection, rather than choosing a specific equation of
state, we shall follow the Poisson-Visser reasoning
\cite{Poisson}, and put quite generally ${\cal P}={\cal
P}(\sigma)$, see (\ref{generalstateequation}). Thus, the potential
given by Eq. (\ref{potential}), may be rewritten as
\begin{equation}
V(a)=\alpha^2 a^2-\frac{M}{\alpha a} -[2\pi a \sigma(a)]^2   \,.
\end{equation}
Linearizing around the stable solution at $a_0$, a second order
expansion of $V(a)$ around $a_0$ provides
\begin{eqnarray}
V(a)&=&V(a_0)+V'(a_0)(a-a_0)
\nonumber    \\
&&+\frac{1}{2}\,V''(a_0)(a-a_0)^2+O \left[(a-a_0)^3 \right]
\label{Taylorexpansion} \,,
\end{eqnarray}
where the prime denotes a derivative with respect to $a$. To
determine $V'(a)$ and $V''(a)$, it is useful to rewrite the
conservation of the surface stress-energy tensor, Eq.
(\ref{conservationofenergy}), as $\sigma 'a=-2(\sigma +{\cal P})$,
taking into account $\sigma '=\dot{\sigma}/\dot{a}$. Defining the
parameter $\eta(\sigma)$ by
\begin{equation}
\eta(\sigma)=d{\cal P}/d\sigma\,, \label{eta}
\end{equation}
and since ${\cal P}(\sigma)'=(d{\cal
P}/d\sigma)\sigma'=\eta\sigma'$, we have $\sigma '+2{\cal
P}'=\sigma '(1+2\eta)$. Thus, $V'(a)$ and $V''(a)$, are given by
\begin{eqnarray}
V'(a)&=&\frac{M}{\alpha a^2}+2\alpha ^2 a+8\pi^2\sigma a\,(\sigma
+2{\cal P})   \label{firstderivativeV}  \,, \\
V''(a)&=&-\frac{2M}{\alpha a^3}+2\alpha ^2 -8\pi^2 \Big[(\sigma
+2{\cal P})^2
       \nonumber     \\
&&+2\sigma (\sigma +{\cal P})(1+2\eta) \Big] \label{2ndderivative}
\,,
\end{eqnarray}
respectively. Evaluated at the static solution, using Eqs.
(\ref{membranesigma})-(\ref{membranepressure}), and taking into
account Eqs. (\ref{sigma+P})-(\ref{sigma+2P}), we find $V(a_0)=0$
and $V'(a_0)=0$, with $V''(a_0)$ given by
\begin{equation}
V''(a_0)=-\alpha^2\,\frac{3M}{(\alpha a_0)^3}\left(\frac{\alpha ^2
a_0^2+\frac{M}{2\alpha a_0}}{\alpha ^2 a_0^2-\frac{M}{\alpha
a_0}}-2\eta_0 \right)  \,, \label{potential3}
\end{equation}
where $\eta_0=\eta (\sigma_0)$. For this, note that Eq.
(\ref{potential3}) may be expressed as
\begin{equation}
V''(a_0)=\alpha^2\,\frac{3M}{(\alpha a_0)^3}\left(\frac{2{\cal
P}_0}{\sigma_0}+1+2\eta_0 \right), \label{potential4}
\end{equation}
by taking into account the surface stresses evaluated at the
static solution $a_0$, given by Eqs.
(\ref{membranesigma})-(\ref{membranepressure}). The potential
$V(a)$, Eq. (\ref{Taylorexpansion}), is reduced to
\begin{equation}
V(a)=\frac{1}{2}\,V''(a_0)(a-a_0)^2+O \left[(a-a_0)^3 \right]  \,,
\end{equation}
so that the equation of motion for the wormhole throat presents
the following form
\begin{equation}
\dot{a}^2=-\frac{1}{2}V''(a_0)(a-a_0)^2+O \left[(a-a_0)^3 \right]
\,, \label{eqmotion}
\end{equation}
to the order of the approximation considered. The solution is
stable if and only if $V''(a_0)>0$. Three cases may be
distinguished, namely, $0<{M}/{(\alpha a_0)^3}<1$ (which
essentially amounts to $M>0$), ${M}/{(\alpha a_0)^3}=0$ (which
essentially amounts to $M=0$), and ${M}/{(\alpha a_0)^3}<0$ (which
essentially amounts to $M<0$). For $0<{M}/{(\alpha a_0)^3}<1$, the
solution is stable if
\begin{equation}
\eta_0  > \frac12\,\left[\frac{(\alpha a_0)^2+ \frac{M}{2\alpha
a_0}} {(\alpha a_0)^2-\frac{M}{\alpha a_0}}\right]\,, \quad
0<\frac{M}{(\alpha a_0)^3}<1\,. \label{inequality1-0}
\end{equation}
For the specific case of ${M}/{(\alpha a_0)^3}=0$ we have a static
and neutrally stable wormhole solution. This can be readily
verified from Eq.  (\ref{potential4}), which implies $V''(a_0)=0$,
and consequently $\dot{a}=0$, from Eq.  (\ref{eqmotion}). Thus any
$\eta_0$ is possible, i.e.,
\begin{equation}
-\infty<\eta_0  <\infty\,,  \quad \frac{M}{(\alpha a_0)^3}=0\,.
\label{inequalitymass0-1}
\end{equation}
Note that this is consistent with the analysis outlined in
Sections \ref{darkenergy} and \ref{gencon}. For ${M}/{(\alpha
a_0)^3}<0$, the stability region is dictated by the following
inequality
\begin{equation}
\eta_0  < \frac12\,\left[\frac{(\alpha a_0)^2-\frac{|M|}{2\alpha
a_0}} {(\alpha a_0)^2+\frac{|M|}{\alpha a_0}}\right]\,, \quad
\frac{M}{(\alpha a_0)^3}<0 \,. \label{inequality2-0}
\end{equation}
The stability regions are plotted in Figure \ref{fig:stable1}.
One can put Eqs. (\ref{inequality1-0})-(\ref{inequality2-0}) in
terms of the matter fields. These yield,
\begin{equation}
\eta_0  > - \left(\frac{{\cal P}_0}{\sigma_0}+\frac12\right)\,,
\quad 0<\frac{M}{(\alpha a_0)^3}<1\,, \label{inequality2}
\end{equation}
\begin{equation}
-\infty<\eta_0  <\infty\,,  \quad \frac{M}{(\alpha a_0)^3}=0\,,
\label{inequalitymass0-2}
\end{equation}
\begin{equation}
\eta_0  < - \left(\frac{{\cal P}_0}{\sigma_0}+\frac12\right)\,,
\quad \frac{M}{(\alpha a_0)^3}<0\,, \label{inequality1}
\end{equation}
respectively.

\subsubsection{Specific equation of state: dark energy}
\label{darkenergy2}

It is also interesting to consider the stability regions, using
the linearization approach, for the specific case of the dark
energy equation of state, ${\cal P}=\omega\sigma$, (see Eq.
(\ref{eos})). By considering ${\cal P}=\omega\sigma$, we note that
for this linear equation of state, we have $\eta_0=\omega$. Now,
using the stability condition $V''(a_0)>0$, from Eqs.
(\ref{inequality1})-(\ref{inequality2}) we have the following
stability conditions. For $M>0$, one has to take into account the
relation between $\omega$ and the mass $M$ given in Eq.
(\ref{Gamma}) (see also Fig. \ref{Gamma-range}), i.e., if $M>0$
then $\omega$ is in the range $\omega<-1$, whereas the range
$\omega>-\frac{1}{4}$ is physically unacceptable since it violates
the no horizon condition. So, for $M>0$, $\omega$ is in the range
$\omega<-1$, and the solutions are unstable. For $M=0$ one has
$\omega=-1$, and the solution is stable. For $M<0$, $\omega$ is in
the range $-1<\omega<-1/4$, and the solutions are stable. Thus, we
finally have
\begin{eqnarray}
-\infty<\omega<-1, \; {\rm i.e.}, \; 0<\frac{M}{(\alpha a_0)^3}<1
\,, {\rm unstable}\,,\\
\omega=-1,\; {\rm i.e.}, \quad\quad \frac{M}{(\alpha a_0)^3}=0
\,,\quad\,{\rm stable}\,,\\
-1<\omega<-\frac{1}{4},\; {\rm i.e.}, \quad\quad \frac{M}{(\alpha
a_0)^3}<0 \,,\quad {\rm stable}\,,
\end{eqnarray}
respectively. So, thin shell wormholes made of phantom energy
($\omega<-1$, $M>0$) are unstable, while thin shell wormholes made
of dark energy fluid ($-1<\omega<-1/4$, $M<0$) or of cosmological
constant fluid ($\omega=-1$, $M=0$) are stable.

\subsection{Comparison between both stability analyses}

Of course both stability analyses have to agree. That this is so
one can deduce by comparing the specific examples considered in
section \ref{darkenergy} and the results given in section
\ref{darkenergy2}.  We have found in section \ref{darkenergy} that
for $\omega=-3/2$ and $M>0$ the solution is unstable, which is
readily confirmed above. On the other hand we have found that for
$\omega=-1/2$ and $M<0$ the solution is stable which is also
verified.

\bigskip

\section{Conclusion}\label{Conclusion}

Using the cut-and-paste technique, we have constructed an elegant
and simple class of static and dynamic plane symmetric wormholes
by surgically grafting together two spacetimes of plane symmetric
vacuum solutions with a negative cosmological constant. We have
further considered a dynamical stability analysis, first, by
considering specific equations of state, and paid special
attention to the dark energy equation of state. Second, a
linearized stability analysis was also explored. Considering
radial perturbations around a static solution, the respective
stability regions were presented, and the specific case of a dark
energy equation of state was also analyzed. It was found that thin
shell wormholes made of a dark energy fluid or of a cosmological
constant fluid are stable, while thin shell wormholes made of
phantom energy are unstable. These plane symmetric wormholes may
be viewed as domain walls connecting different universes. The
construction of these wormholes does not alter the topology of the
background spacetime, i.e., spacetime is not multiply-connected,
so that, in general, these wormhole solutions do not allow time
travel.

\bigskip

\begin{acknowledgments}

\noindent FSNL was funded by Funda\c{c}\~{a}o para a Ci\^{e}ncia e
Tecnologia (FCT)--Portugal, through the grant SFRH/BPD/26269/2006.
This work was also partially funded through the  FCT project
POCI/FP/63943/2005.

\end{acknowledgments}

\end{document}